# NMR investigation of the pressure induced Mott transition to superconductivity in $Cs_3C_{60}$ isomeric compounds

**H Alloul[1], Y Ihara[1], T Mito[1], P Wzietek[1], M Aramini[2], D Pontiroli[2] and M Ricco[2]**

[1] Laboratoire de Physique des Solides, CNRS UMR 8502, Université Paris-Sud 11 91405 Orsay (France)

[2] Dipartimento di Fisica, Università di Parma Via U.P. Usberti 7/a, 43100 Parma (Italy)

Email: alloul@lps.u-psud.fr

**Abstract**. The discovery in 1991 of high temperature superconductivity (SC) in $A_3C_{60}$ compounds, where $A$ is an alkali ion, has been initially ascribed to a BCS mechanism, with a weak incidence of electron correlations. However various experimental evidences taken for compounds with distinct alkali content established the interplay of strong correlations and Jahn Teller distortions of the $C_{60}$ ball. The importance of electronic correlations even in $A_3C_{60}$ has been highlighted by the recent discovery of two expanded fulleride $Cs_3C_{60}$ isomeric phases that are Mott insulators at ambient pressure. Both phases undergo a pressure induced first order Mott transition to SC with a ($p, T$) phase diagram displaying a dome shaped SC, a common situation encountered nowadays in correlated electron systems. NMR experiments allowed us to establish that the bipartite A15 phase displays Néel order at 47K, while magnetic freezing only occurs at lower temperature in the fcc phase. NMR data do permit us to conclude that well above the critical pressure, the singlet superconductivity found for light alkalis is recovered. However deviations from BCS expectations linked with electronic correlations are found near the Mott transition. So, although SC involves an electron-phonon mechanism, correlations have a significant incidence on the electronic properties, as had been anticipated from DMFT calculations.

## 1. Introduction

In fullerides $A_nC_{60}$, the charge transferred from the alkali atoms ($A$) to the fullerene molecules is nearly complete, and is expected to participate in delocalized metallic states. However the weak transfer integrals between $C_{60}$ molecules imply electronic structures with small bandwidths [1], so that electronic correlations should be important [2]. Furthermore adding electrons on the $C_{60}$ balls cannot be done without taking into consideration the coupling of the molecular electronic states with the inner vibration modes of the molecule, which induce Jahn-Teller effects [3]. This yields the observation of a series of effects associated with molecular physics and strong correlations in the $A_nC_{60}$ compounds.

However, from the normal state and superconducting properties of $A_3C_{60}$ compounds one considered initially that those were weakly correlated metals displaying superconducting BCS behaviour [4]. The latter was still thought somewhat original as the pairing appears mediated by local on ball optical phonon modes.

While this convinced many researchers that electronic correlations were not important in the $A_3C_{60}$, further studies of various $A_nC_{60}$ allowed to evidence insulating states with large optical gaps (~0.5eV) in the even electron compounds $A_4C_{60}$ [5-6] and $Na_2C_{60}$ [7] which cannot be explained by a simple progressive band filling of the lowest unoccupied $C_{60}$ six-fold degenerate $t_{1u}$ molecular level.

The absence of magnetism in the ground state and the small spin gap observed by NMR spin lattice relaxation data [7] could only be ascribed to the simultaneous influence of strong electron correlations and Jahn-Teller Distortions (JTD) of the $C_{60}$ ball, which localizes the electrons on the balls in a low spin state [8-9].Those energetically favour evenly charged $C_{60}$ molecules, hence their qualification as Mott-Jahn-Teller insulators [10].

As for the $A_3C_{60}$, recent efforts to study them in expanded states confirm that they are often magnetic [11]. Furthermore, as will be recalled here, the large ionic radius of Cs permits to stabilize two expanded $Cs_3C_{60}$ phases which are magnetic and insulating at ambient pressure and exhibit a Mott transition to a metallic and SC state when the distance between $C_{60}$ is decreased by an applied pressure[12-13]. Those results confirm anticipations from calculations of the ground state energies assuming strong correlations and Jahn -Teller effects [14].

We shall recall first in section 2 the usual attempts to explain SC in $A_3C_{60}$ within the BCS scenario and shall discuss then in section 3 the expected incidence of correlations on this picture. In section 4 the NMR data on the two Mott phases of $Cs_3C_{60}$ will be reported. The phase diagram and the electronic properties of the SC state will be described in section 5. The NMR data on the electronic properties of the metallic state near the Mott transition will be displayed in section 6, and will allow us to evidence the recovery of a metallic state upon thermal expansion of the lattice at high $T$.

## 2. BCS Superconductivity in $A_3C_{60}$?

Contrary to the evenly charged fullerides, the $A_3C_{60}$ compounds are usually metallic and present a regular Drude peak in their optical conductivity [15]. Also the normal state electronic susceptibility, as measured by the electron spin resonance (ESR) intensity is practically $T$ independent [16], that is Pauli like, as confirmed as well by NMR Knight shift and $T_1T$ data [17]. Quantitatively the effective mass enhancement is $m^*/m_0 \sim 3$, so that one is tempted to conclude that these compounds are not far from regular Fermi liquids [4, 18].

In the superconducting state, a jump of the specific heat is detected and increases with $T_c$ as expected for a BCS transition [19]. Similarly, the SC gap $2\Delta$ scales with $T_c$, with a $2\Delta/k_BT_c$ ratio not far from 3.5, value expected in the theoretical BCS scenario [4]. NMR in the superconducting state shows up the decrease of Knight shift, that is the loss of spin susceptibility, expected for singlet pairing [20-21]. Finally Hebel-Slichter coherence peaks in both the NMR $(T_1)^{-1}$ [20-21] and the muon spin relaxation rate [22] have been observed below $T_c$ in weak applied fields, which also points towards *conventional s-wave symmetry of the order parameter.*

But, among the experimental observations, the most emphasized one has been the variation of $T_c$ with the chemical nature of the inserted alkali, which permits to change the inter-ball distance [4]. One could monitor, as shown in figure 1, a continuous decrease of $T_c$ with lattice compression in fcc $A_3C_{60}$.

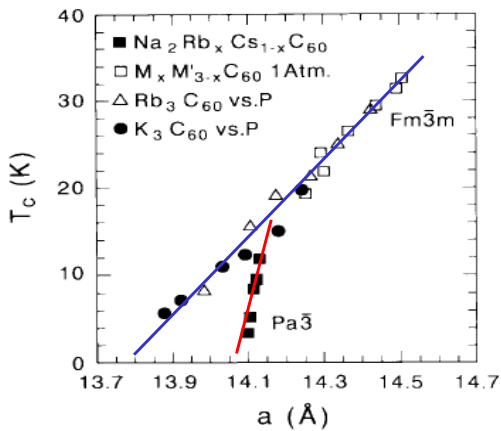

**Fig. 1.** $T_c$ variation versus fcc lattice parameter obtained by chemical exchange of the alkali (M /M') or by applying a pressure which leads to a decrease of DOS and of $T_c$. The samples with cubic *Pa*-3 structure do not follow the same trend as those with *Fm-3m* structure (adapted from [23]).

There, the detected $T_c$ variation is the same for chemical [24] and pressure changes [25] of the lattice parameter. Moreover, the absence of alkali mass isotope effect has been initially assigned to a BCS weak coupling relation for $T_c$

$$k_B T_c = \hbar\omega_D \exp[-1/VN(E_F)], \qquad (1)$$

in which the density of states $N(E_F)$ appears to be the single driving parameter. This has led to consider an on ball electron-phonon coupling $V$, the active phonons in $\omega_D$ being then the high energy optical phonons (Hg), which are responsible for the JT effects [4] [26-28].

However, it has been pointed out that, as displayed in figure 1, this relation does not extend to samples of $Na_2AC_{60}$, which order in the cubic $Pa3$ structure [29-30], though a similar law might hold with a larger slope [23]. This difference has raised questions about the possible incidence of the structure, of the nature of the alkali and of the merohedral disorder on the density of states [31] which has to be considered in equ.(1). Actually, it has been shown by ARPES that the arrangement of the molecules on deposited single layer fulleride films at the half filled composition does have a dominant influence on the electronic properties of the films [32].

### 3. Incidence of JTD and electronic correlations

Considering these results one is led to question whether the correlation effects and JTD which induce the Mott insulating state of the evenly charged compounds do play any role in the electronic properties of these $A_3C_{60}$ compounds. We have seen that the Mott JT states for even $n$ are already not so far from a metallic state. Does that mean that odd parity of $n$ would indirectly favor hopping rather than localization?

To answer that question, one might consider the energy cost of transferring one electron on a nearby site, that is transforming $2C_{60}^{-n}$ into $C_{60}^{-(n+1)}$ and $C_{60}^{-(n-1)}$. If $E(n)$ is the molecular energy of a ball with $n$ electrons, this energy cost is given by

$$U_{eff} = E(n+1) + E(n-1) - 2E(n). \qquad (2)$$

One has been led to conclude that the contribution of the JTD to $U_{eff}$ will be positive for even $n$, and negative for odd $n$. The JTD stabilisation of evenly charged $C_{60}$ therefore adds to the on ball coulomb repulsion for even $n$ and reduces it for odd $n$ [33-34]. This explains the occurrence of large correlation effects and of the JTD Mott insulating state in the evenly charged compounds. But conversely this can be taken as an argument for smaller overall electronic correlations in the $A_3C_{60}$ compounds. However the magnitude of $U_{eff} < 0.5$ eV, as compared to $U \sim 1$ eV, would hardly justify a total loss of electronic correlations [4].

Various efforts have then been undertaken to estimate the incidence of electronic correlations on electron phonon superconductivity in the case of these $A_3C_{60}$. Usually, the Coulomb electronic interactions are taken into account within the McMillan extension of BCS through an empirical parameter $\mu^*$

$$k_B T_c = \hbar\omega_D \exp[-1/(\lambda-\mu^*)], \qquad (3)$$

where $\lambda = VN(E_F)$. This equation applies when $\hbar\omega_D \ll E_F$ while here the phonons, the JTD and the Fermi energies have the same order of magnitude, which does not allow a standard use of this approach[1]. More recent calculations done within Dynamical Mean Field Theory (DMFT), suggest that in this specific case of Jahn-Teller on ball phonons, the electronic correlations would not be very detrimental to phonon driven superconductivity [35].

---

[1] Using the more accurate Migdal-Eliashberg equation does not change this conclusion, see [4].

In any case a different school of thought has been suggesting for long that electron-electron correlations might even drive superconductivity in such compounds [2-3]. It has even been proposed that $A_nC_{60}$ should be Mott insulators and are only metallic due to a non stoichiometry of the alkalis [36]. Although our result for the charge segregation in the cubic quenched phase of $CsC_{60}$ [37] could be supportive of this possibility, no experimental evidence along this line could be found so far in $A_nC_{60}$. It has rather been seen that $T_c$ peaks at $n=3$ if one varies the alkali doping [38].

To better qualify the proximity to a Mott insulating state, attempts to produce expanded fulleride compounds have been done for long. A successful route has been to insert $NH_3$ neutral molecules to expand the $A_3C_{60}$ lattice. It was found that $(NH_3)K_3C_{60}$ is insulating at ambient pressure [39], and becomes SC with $T_c$ up to 28K under pressure [40]. Furthermore at 1bar it was shown to be an AF [11]. However this insertion of $NH_3$ induces an orthorhombic distortion of the lattice which is retained, though modified, under applied pressure [41]. So, although this system undergoes a Mott transition to a metallic state, this occurs in an electronic structure where the $t_{1u}$ level degeneracy has been lifted by the specific spatial order of the $C_{60}$ balls. Let us recall then that, for a single orbital case the Mott transition occurs for $U/W=1$, while for $N$ degenerate orbitals the critical value $U_c$ is expected to be larger [33], typically $U_c=N^{1/2}W$. Therefore $U_c$ is reduced in the case of $(NH_3)K_3C_{60}$ with respect to that of $A_3C_{60}$, for which the $t_{1u}$ orbital degeneracy is preserved. Though this complicates any direct comparison between these two cases, these results establish that $A_3C_{60}$ are not far from a Mott transition, and hence that electronic correlations cannot be neglected.

**4. Mott insulating state in $Cs_3C_{60}$ isomeric compounds**

A much simpler way to expand the $C_{60}$ fcc lattice has been of course to attempt the insertion of Cs, the largest alkali ion, by the synthesis procedure used for the other $A_3C_{60}$. But, the bct $Cs_4C_{60}$ and the polymer phase $Cs_1C_{60}$ being stable at room $T$, the $Cs_3C_{60}$ composition is found metastable. In a mixed composition sample it was found that a $Cs_3C_{60}$ phase becomes SC at high pressure with $T_c \sim 40$ K [42]. A distinct chemical route finally recently permitted to produce samples containing the fcc-$Cs_3C_{60}$, mixed with an A15 isomeric phase. The latter has been found insulating at ambient pressure, and becomes SC under pressure, exhibiting a SC dome with a maximum $T_c$ of 38K at about 10 kbar [12].

In similar mixed phase samples, NMR experiments allowed us to separate the specific $^{133}Cs$ spectra of the two isomers [13]. As displayed in figure 2, the A15 phase has a single Cs site with non cubic local symmetry, hence its seven lines quadrupole split spectrum ($^{133}Cs$ has a nuclear spin I=7/2). The fcc-$Cs_3C_{60}$ NMR spectra exhibit the usual features found for all other fcc-$A_3C_{60}$, displaying the signals of the orthorhombic (O) site and of two tetrahedral sites (T, T') [43]. Those are associated with the merohedral disorder [44] of the orientation of the $C_{60}$ balls, which differentiates two local orientation orderings of the $C_{60}$ around the alkali [45].

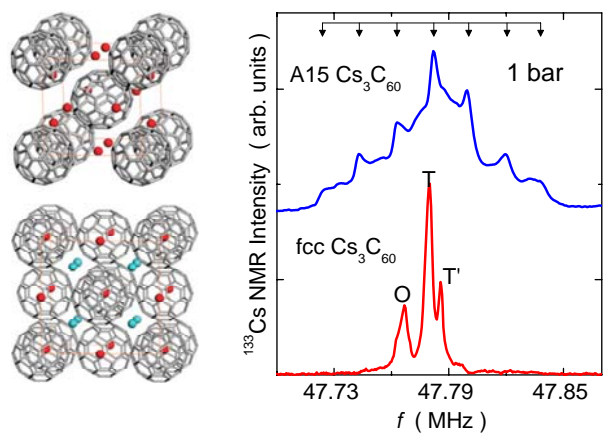

**Fig. 2.** The unit cells of the A15 and fcc isomeric phases of $Cs_3C_{60}$ are displayed on the left [12]. Their $^{133}Cs$ NMR spectra [13] taken at 1bar and $T$=300 K are reported on the right. The specific features of the spectra are discussed in the text.

This possibility to select the signals of the two phases allowed us to demonstrate that both are in a paramagnetic insulating state at $p=1$ bar, and display a Mott transition at distinct applied pressures[13]. We could further study their magnetic behaviour at ambient pressure using data on the $^{13}$C nuclear spins, which are directly coupled to the electron spins of the $t_{1u}$ molecular orbitals. The coupling being dipolar [21], it induces a characteristic anisotropic broadening of the $^{13}$C NMR which allowed us to monitor the $C_{60}^{3-}$ magnetism.

As can be seen in figure 3 this anisotropic shift $K_{ax}$ evidences a Curie–Weiss paramagnetism of the $C_{60}^{3-}$, which does not differ for the two isomers above 100K [13]. These data are compatible with a low spin $S=1/2$ effective moment as expected from figure 3 for a JT stabilized state. Furthermore the structural difference between the two isomers results in distinct magnetic ground states at $p=1$ bar. The bipartite A15 structure displays a Néel order below $T_N=47$ K [12,13,46], while the frustration of the fcc lattice results in a spin freezing only below 10 K for the standard fcc-$Cs_3C_{60}$ [13,47].

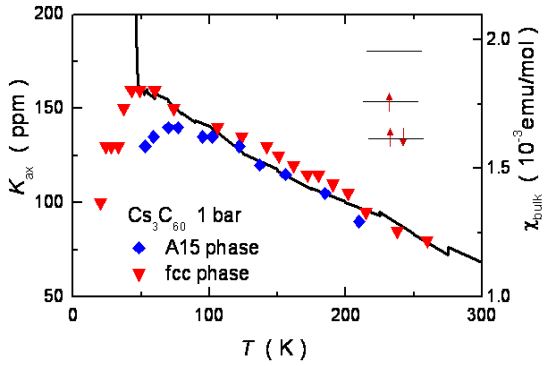

**Fig. 3.** Paramagnetic behaviour detected by SQUID data (full line, right scale) and from the anisotropic shift $K_{ax}$ of the $^{13}$C NMR in the two isomeric $Cs_3C_{60}$ phases [13]. No significant difference can be detected above 100 K and the effective moment corresponds to the low spin JTD state sketched in insert. The magnetization increase below 47 K is that of the A15 phase Néel state.

## 5. SC state induced under applied pressure and low *T* phase diagram

Under an applied pressure, macroscopic diamagnetism is found to appear abruptly in both phases though at a slightly lower critical pressure $p_c$ for the fcc phase than for the A15. From NMR data in the A15 phase we could evidence that this appearance of SC is concomitant with an abrupt loss of magnetism, which definitely points toward a first order transition. The phase diagrams slightly differ for the two $Cs_3C_{60}$ isomers, but as shown in figure 4, they roughly merge together [13,47] if plotted versus $V_{C60}$, the unit volume per $C_{60}$ ball. There, the $T_c$ data can be scaled as well with that for the other known fcc-$A_3C_{60}$, and it has been evidenced that a similar maximum of $T_c$ versus $V_{C60}$ applies for the two structures [13,47,48].

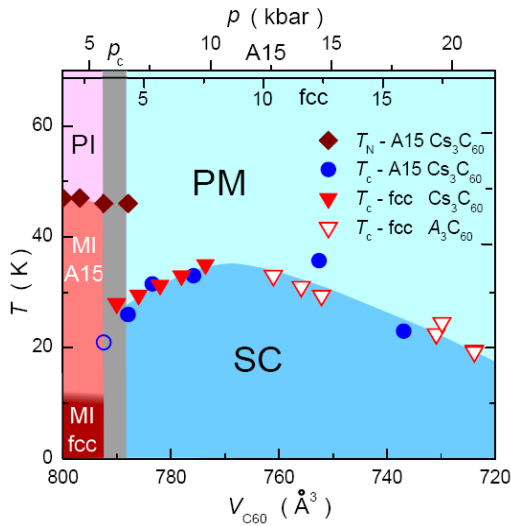

**Fig. 4.** Phase diagram representing $T_c$ versus the volume $V_{C60}$ per $C_{60}$ ball, and the Mott transition at $p_c$ (hatched bar) for the two isomer phases [47]. The corresponding pressure scales are shown on the upper scale. Data for the other known fcc-$A_3C_{60}$ are reported as empty symbols. The transition to a magnetic insulating (MI) phase occurs only below 10 K in fcc-$Cs_3C_{60}$.

We could probe the electronic excitations by taking $^{133}$Cs NMR spin lattice relaxation data. Well above $p_c$, drops of $(T_1T)^{-1}$, which evidence the opening of the superconducting gap below $T_c$, were detected for the two phases [13]. So the electronic properties are found to behave as for the non expanded fcc-phases, with a singlet state s-wave SC state. This ensemble of results reveal that the SC state does not depend significantly on the actual structure, except through the modification of the inter-ball distance and transfer integrals between neighbouring $C_{60}$.

As seen in figure 5, we could evidence in the A15 phase that such a drop of $(T_1T)^{-1}$ already appears just above $p_c$, together with a diamagnetic shift of the NMR signal [49] which confirms that *bulk SC* occurs immediately after the first order Mott transition.

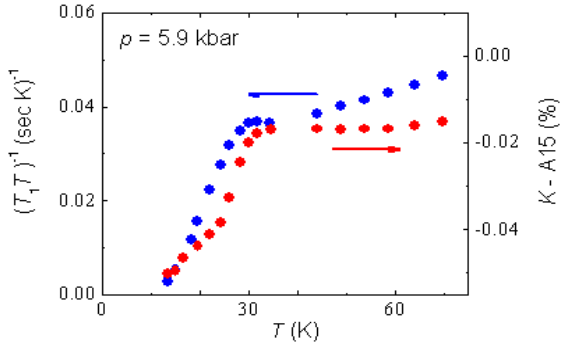

**Fig. 5.** $^{133}$Cs NMR $(T_1T)^{-1}$ (left scale) and shift data (right scale) taken at a pressure just above $p_c$ ~ 5.1 kbar for a nearly pure A15 sample. Bulk SC is evidenced to occur below about 30K at this pressure.

Above the magnetic and SC ground states, the x-ray and NMR spectra did not display any significant change with increasing pressure through the paramagnetic insulator to metal transition, for both isomers. This absence of significant structural modifications at $p_c$ establishes then that the evolution with $p$ is dominated by electronic degrees of freedom.

## 6. Normal state electronic properties and high $T$ insulating properties

A great advantage of NMR $(T_1T)^{-1}$ data is that it permits us to probe the evolution of the electronic properties from the paramagnetic insulator to the paramagnetic metallic state. This is illustrated in figure 6 by a series of measurements performed in the fcc phase on the $^{133}$Cs NMR of the T' site [47].

One can see that, in the Mott state, below the pressure $p_c$ of the Mott transition, $(T_1T)^{-1}$ increases with decreasing $T$ as expected for a dense paramagnet, this being true for both phases. In fact in this regime $(T_1)^{-1}$ is nearly $T$ independent well above the magnetically ordered state, its value being mainly determined by the exchange coupling $J$ between the moments localized on the $C_{60}$ balls [13,46,47].

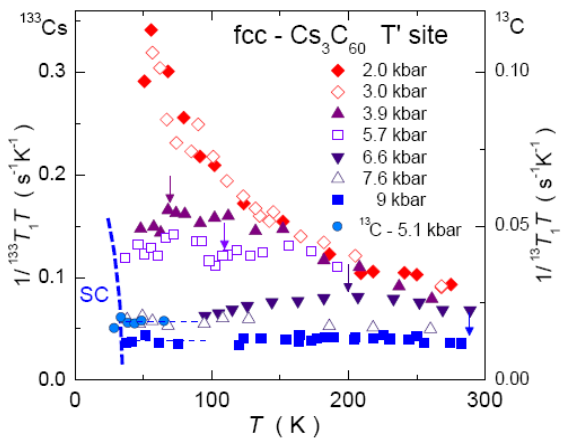

**Fig. 6.** $^{133}$Cs $(T_1T)^{-1}$ data taken at increasing pressures on the T' site of fcc-$Cs_3C_{60}$. The low $T$ Mott transition occurs at $p_c$~3.9 kbar. Notice the sharp decrease of $(T_1T)^{-1}$ at the MIT, and then its regular decrease with increasing pressure in the metallic state. The recovery of an insulating state upon expansion of the lattice at high $T$ can be seen from the progressive evolution of $(T_1T)^{-1}$ toward that of the low pressure paramagnetic insulator. From [47]

At high pressures above 7 kbar a nearly $T$ independent $(T_1T)^{-1}$ is recovered, as indeed seen in most other non expanded fcc phases, and as expected for the Korringa nuclear relaxation of a metallic state

$$(T_1T)^{-1} = (4\pi k_B / \hbar)\, A_{hf}^2\, N(E_F)^2, \qquad (4)$$

where $A_{hf}$ is the hyperfine coupling of the $^{133}$Cs nuclear spin with the electronic density of state $N(E_F)$ taken at the Fermi level.

However for intermediate pressures the variation of $(T_1T)^{-1}$ is quite unexpected as it goes through a maximum, pointed out by arrows in figure 6. One can notice that it shifts towards higher $T$ at increasing pressures, while $(T_1T)^{-1}$ progressively approaches the value found in the insulating state.

Let us consider first the behaviour found just above the SC state at about 50 K. We can see in figure 6 the regular increase of $(T_1T)^{-1}$ with decreasing pressure, which would suggest from equ (4) a regular increase of the density of state $N(E_F)$ until the Mott transition at $p_c$. But such an increase of $N(E_F)$, would correspond in the BCS formalism of Equ(1) to an increase of $T_c$, while in this range of pressures, $T_c$ decreases when $p_c$ is approached from above (this corresponds to the decreasing part on the left of the dome in figure 4). Consequently equations (4) or (1), or both, do not apply [50]. In other words the magnetic excitations detected above the Mott transition have some specific incidence on the SC condensation energy.

The second important observation illustrated in figure 6 is the recovery of the magnetic excitations of the paramagnetic insulating state which occurs at high temperature for any fixed pressure. The fulleride compounds being highly compressible, they also of course expand with increasing $T$, so that the Mott transition is crossed as well at high $T$ and the insulating state is recovered in this expanded state. In other words the critical pressure $p_c$ increases with increasing $T$. Let us point out that such a transition toward an insulating state had been suggested about a decade ago to explain the constant value of $(T_1)^{-1}$ found above room $T$ in the cubic $Na_2CsC_{60}$ compounds [51]. Finally, as discussed in [47] the transition line $p_c(T)$ seems to reach around room $T$ a critical point similar to that found for the liquid gas transition. At higher temperatures a continuous evolution from the paramagnetic metallic state to the paramagnetic insulating state should occur. This behaviour appears quite analogous to that observed for the 2D Mott transition in organic compounds [52].

## 7. Discussion

We have recalled here that electronic correlations and on ball Jahn-Teller effects have significant incidence on the physical properties of $A_nC_{60}$ compounds. Those help to establish a robust insulating state for evenly doped compounds. While electronic correlations were not expected to be dominant in the $n=3$ cubic compounds, we have definitely shown here that those remain quite close to a Mott transition.

The expanded $Cs_3C_{60}$ compounds are unique inasmuch as they are Mott insulators at $p=1$ bar and can be driven through the Mott transition by an applied pressure. The Mott state is found to correspond to low spin as could be expected for a Jahn-Teller distorted ground state. This has been confirmed recently by infrared experiments [53] which do evidence a dynamic loss of symmetry of the $C_{60}$ balls. The phase diagram displays a SC dome, reminiscent of that found in many other correlated electron systems, with a $T_c$ maximum at a pressure somewhat above that of the first order Mott insulator state boundary.

In the SC state, the NMR experiments have established that the pure BCS equation or its Migdal-Eliashberg extension do not yield even a qualitative explanation of the deviations found near the Mott transition, where $T_c$ decreases with decreasing pressure. There the normal state electronic properties deviate from Fermi liquid behaviour as spin fluctuations in the metallic state become prominent before the system switches into the Mott state [47]. Also, the orientational disorder of the $C_{60}$ in fcc-$Cs_3C_{60}$ does not appear detrimental to the $s$-wave SC, even near the Mott transition, as the $T_c$ dome does not differ from that of the A15 ordered isomer. This opposes the case of cuprates for which both normal state and $d$-wave SC are very sensitive to disorder [54-55].

On the theory side, serious efforts to include JT effects and correlations using DMFT have allowed to evidence that the Coulomb repulsion is at least not that detrimental to this on ball pairing [35]. It is important to recall here that ground state energy calculations done with DMFT have even allowed the anticipation [56] of the phase diagram with a SC dome near the Mott transition [57]. This is a strong indication that detailed theoretical understanding of the properties of this series of materials is within reach. Further developments should then help to decide whether the large $T_c$ values in $A_3C_{60}$ compounds do result from a fundamental cooperation between correlations and electron-phonon interactions, and to determine above which pressure or bandwidth the BCS limit would be recovered.

Finally, we propose here that $Cs_3C_{60}$ is a rather good model system to study a 3D Mott transition. Indeed the latter is not controlled by doping, but is only driven by the inter-ball distance, that is the $t_{1u}$ bandwidth. In the *(p,T)* range probed in [47] the qualitative behaviour detected appears quite similar to that expected for a single orbital Mott transition [58] although the *S*=1/2 ground sate has orbital degeneracy. Extensive measurements of other spectral and thermodynamic properties, possibly on larger *T* ranges, are required to complete the experimental insight on this Mott transition.

**Acknowledgments**
The authors acknowledges particularly V. Brouet who has been the major actor in the work on the non SC compounds and shows a continuing interest on the matter. We thank as well L. Forró and his group, with whom collaborations have been extending over many years.

We also warmly thank E. Tosatti, M. Fabrizio and M Capone with whom we have had over the years many fruitful discussions about the theoretical aspects of the electronic structure of the fulleride compounds.